\documentclass[a4paper]{cs13_cranmer}
\usepackage{epsfig}
\usepackage{cranm_timesv}

\begin{document}

\title{New insights into Solar Wind Physics from SOHO}

\author{S. R. Cranmer}
\institute{Harvard-Smithsonian Center for Astrophysics,
Cambridge, MA 02138, USA}

\maketitle 

\begin{abstract}

\noindent
The Solar and Heliospheric Observatory ({\em{SOHO}}) was
launched in December 1995 with a suite of instruments designed
to answer long-standing questions about the Sun's internal
structure, its extensive outer atmosphere, and the solar wind.
This paper reviews the new understanding of the physical
processes responsible for the solar wind that have come from the
past 8 years of {\em SOHO} observations, analysis, and
theoretical work.
For example, the UVCS instrument on {\em SOHO} has revealed the
acceleration region of the fast solar wind to be far from simple
thermal equilibrium.
Evidence for preferential acceleration of ions, 100 million K
ion temperatures, and marked departures from Maxwellian velocity
distributions all point to specific types of collisionless
heating processes. 
The slow solar wind, typically associated with bright helmet
streamers, has been found to share some of the nonthermal
characteristics of the fast wind.
Abundance measurements from spectroscopy and visible-light
coronagraphic movies from LASCO have led to a better census
of the plasma components making up the slow wind.
The origins of the solar wind in the photosphere and
chromosphere have been better elucidated with disk spectroscopy
from the SUMER and CDS instruments.
Finally, the impact of the solar wind on spacecraft systems,
ground-based technology, and astronauts has been greatly aided
by having continuous solar observations at the Earth-Sun L1
point, and {\em SOHO} has set a strong precedent for future
studies of space weather.


\end{abstract}

\section{Introduction}

If the Sun is a benchmark for stellar astrophysics, then the
solar wind is even more of a necessary reference for the
study of stellar winds.
The Sun is special because of its proximity;
the solar wind is {\em unique} because we are immersed in it
and its plasma can be accessed directly by space probes.
Despite all that has been learned by the {\em in situ} 
detection of particles and fields, though, we have learned the
most about how the solar wind is produced from good,
old-fashioned astronomical imaging and spectroscopy.
This paper summarizes the most recent results of this kind
from the past decade of observations with the {\em SOHO}
spacecraft.
(A top-ten list of reasons ``Why stellar astronomers should
be interested in the Sun'' is given in an article with that
title by Schmelz 2003).

\section{Brief History (pre-SOHO)}

\begin{figure}[t]
  \begin{center}
    \epsfig{file=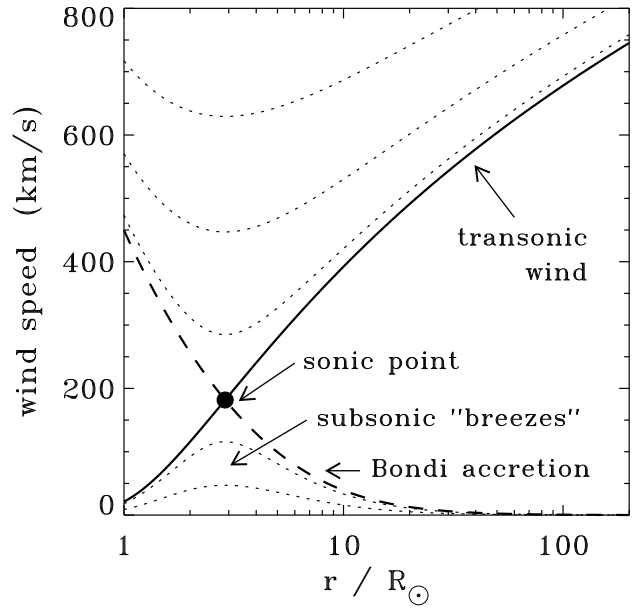, width=8.4cm}
  \end{center}
\caption{Illustration of the analytic solution topology of the
isothermal, time-steady flow equation in spherical geometry,
showing Parker's (1958) solar wind solution (solid line),
Bondi's (1952) accretion solution (dashed line), and other
unphysical solutions (dotted lines).
\label{fig1}}
\end{figure}

Sightings of the solar corona and the shimmering aurora
(i.e., the beginning and end points of the solar
wind that encounters the Earth) go back into antiquity.
The first scientific understanding of the outer solar
atmosphere came as spectroscopy began to be applied
to heavenly light in the late 19th century.
In 1869, Harkness and Young first observed the 5303 {\sc\aa}
coronal green line during a total eclipse.
Soon after, in the very first issue of {\em Nature,}
J.\  Norman Lockyer (the discoverer of helium)
reported that these observations were
``...{\em{bizarre}} and puzzling to the last degree!''
The chemical element responsible for the green line went
unidentified for 70 years, after which Grotrian and Edl\'{e}n
applied new advances in atomic physics to show that several
of the coronal emission lines are produced by very high
ionization stages of iron, calcium, and nickel.
The puzzle then shifted to explaining what causes the outer
solar atmosphere to be heated to temperatures of more than
10$^6$ K.
Despite many proposed theoretical processes, this
``coronal heating problem'' is still with us today because
there are no clear observational constraints that can
determine which if any of the competing mechanisms is
dominant.

Knowledge about an outflow of particles from the Sun was
starting to coalesce at the beginning of the 20th century.
Researchers came to notice strong correlations between
sunspot activity, geomagnetic storms, auroral appearances,
and motions in comet tails.
Parker (1958, 1963) combined these empirical clues with the
earlier discovery of a hot corona and postulated a
theoretical model of a steady-state outward expansion of
gas from the solar surface.
Figure 1 shows analytic solutions to Parker's isothermal
solar wind equation, as well as other solutions including
the case of spherical accretion found earlier by Bondi (1952).
Traditionally this equation was believed to be unsolvable
explicitly for the outflow speed, but Cranmer (2004) showed
how a new transcendental function (the Lambert $W$ function)
can be used to produce exact closed-form solutions of
this equation.

Parker's key insight was that the high temperature of the
corona provides enough energy per particle to overcome
gravity and produce a natural transition from a subsonic
(bound, negative total energy) state near the Sun to a
supersonic (outflowing, positive total energy) state
in interplanetary space.
An initial controversy over whether Parker's ``transonic''
solution or Chamberlain's always-subsonic ``breeze''
solutions were
\linebreak[4]
most physically relevant was dispelled
when {\em Mariner 2} confirmed the existence of a
continuous, supersonic solar wind in interplanetary
space (Snyder and Neugebauer 1964; see also Neugebauer
1997).
Also, Velli (2001) showed that the subsonic breeze
solutions are formally unstable, and that the truly stable
solution for the solar wind case is the one that goes
through the sonic point.

In the years since {\em Mariner 2,} many other deep-space
missions have added to our understanding of the solar wind.
The turbulent inner heliosphere was probed by the two
{\em Helios} spacecraft, which measured particle and field
properties between
\linebreak[4]
0.29 and 1 AU (Marsch 1991).
In the 1990s, {\em Ulysses} became the first probe to venture
far from the ecliptic plane and soar over the solar poles to
measure the solar wind in three dimensions (Marsden 2001).
The {\em Voyager} probes are still sending back data on
the outer reaches of the solar wind, and one of them may
have passed through the termination shock separating the
heliosphere from the interstellar medium (e.g.,
Krimigis et al.\  2003).

It has not been possible to send space probes closer to the
Sun than about the orbit of Mercury, so our understanding
of the corona is limited to remote-sensing observations.
Ultraviolet and X-ray studies of the ``lower'' corona (i.e.,
within about 0.1 to 0.3 $R_{\odot}$ from the surface) began
in the 1960s and kept improving in spatial resolution
from {\em Skylab} in the 1970s (Vaiana 1976)
to {\em Yohkoh} in the 1990s (Martens and Cauffman 2002).
Until the 20th century, total solar eclipses were the only
means of seeing any hint of the dim ``extended'' corona
(heights above $\sim$0.3 $R_{\odot}$)
where most of the solar wind's acceleration occurs.
However, with the invention of the disk-occulting
coronagraph by Lyot in 1930 and the
development of rocket-borne ultraviolet coronagraph
spectrometers in the 1970s (Kohl et al.\  1978;
Withbroe et al.\  1982), a continuous detailed
exploration of coronal plasma physics became possible.

\section{The SOHO Mission}

The {\em Solar and Heliospheric Observatory} ({\em{SOHO}})
is the most extensive space mission ever dedicated to the
study of the Sun and its surrounding environment.
{\em SOHO} resulted from international collaborations
between ESA and NASA dating back to the early 1980s (see,
e.g., Huber et al.\  1996) and became a cornerstone mission
in the International Solar-Terrestrial Physics (ISTP)
program.
The {\em SOHO} spacecraft was launched on December 2, 1995
and entered a halo orbit around the Earth-Sun L1 point
two months later.
This orbit allows an uninterrupted view of the Sun, essential
for helioseismology, but the distance (about 4 Earth-Moon
distances) also puts limits on the amount of data telemetry
that can be received.
{\em SOHO} hosts 12 instruments that study the solar interior,
solar atmosphere, particles and fields in the solar wind, and
the distant heliosphere.
Early results from the first year of operations were
presented by Fleck and \v{S}vestka (1997), and a more
up-to-date summary of the mission---along with details
about how to access data and analysis software---is given
by Domingo (2002).

This paper presents results mainly from the 5 instruments
designed to observe the hot, outer atmosphere of the Sun
(excluding the photosphere) where the solar wind is
accelerated.
These instruments are listed below in alphabetic order.
\begin{enumerate}
\item
{\bf CDS}
(Coronal Diagnostic Spectrometer)
is a pair of extreme ultraviolet spectrometers that view
the solar disk and low off-limb corona in the wavelength
range 150--785 {\sc\aa} with spectral resolution
$\lambda / \Delta\lambda \sim 700$--4500, and with
2--$3''$ spatial resolution
(Harrison et al.\  1995).
\item
{\bf EIT}
(Extreme-ultraviolet Imaging Telescope)
is a full-disk imager with $5''$ spatial resolution
that obtains nar\-row-band\-pass images of the Sun at
384, 171, 195, and 284 {\sc\aa}
(Delaboudini\`{e}re et al.\  1995).
EIT images and movies are probably the most reproduced
data products from {\em SOHO} (see, e.g., the covers of
many popular magazines).
\item
{\bf LASCO}
(Large Angle Spectroscopic Coronagraph) is a
\linebreak[4]
package of 3 visible-light coronagraphs with
overlapping annular fields of view
(Brueckner et al.\  1995).
The C1 coronagraph observes from 1.1 to 3 $R_{\odot}$
with 5 Fabry-Perot filter bandpasses.
The C2 and C3 coronagraphs observe the radii
2--6 $R_{\odot}$ and 4--30 $R_{\odot}$, respectively,
in either linearly polarized or unpolarized light.
C1 was fully operational from launch until the 3-month
{\em SOHO} interruption in June 1998.
\item
{\bf SUMER}
(Solar Ultraviolet Measurements of Emitted Radiation)
is an ultraviolet spectrometer that observes the
solar disk and low corona in the wavelength range
330--1610 {\sc\aa} with a spectral resolution of
about 12000, and with $\sim$1.5$''$ spatial resolution
(Wilhelm et al.\  1995).
SUMER observations have been constrained recently by the
need to conserve the remaining count potential of the
detectors.
\item
{\bf UVCS}
(Ultraviolet Coronagraph Spectrometer)
is a combination of an ultraviolet spectrometer and a
linearly occulted coronagraph that observes a
2.5 $R_{\odot}$ long swath of the extended corona,
oriented tangentially to the solar radius, at
heliocentric heights ranging between 1.3 and 12 $R_{\odot}$
(Kohl et al.\  1995).
The spectrometer slit can be rotated around the Sun.
UVCS observes the wavelength range 470--1360 {\sc\aa}
with $\sim$10$^4$ spectral resolution and $7''$
spatial resolution.
\end{enumerate}

\section{SOHO Solar Wind Results}

The bulk of this paper describes results from the above
{\em SOHO} instruments (and associated theoretical work)
concerning the physics of solar wind acceleration and heating.
Figure 2 is an illustrative summary of the topics covered
by the following subsections.

\begin{figure}[t]
  \begin{center}
    \epsfig{file=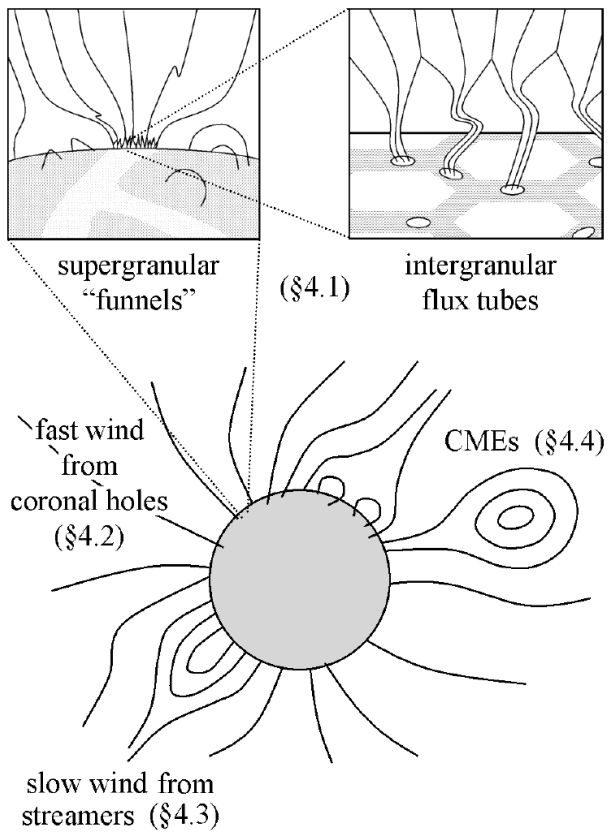, width=8.0cm}
  \end{center}
\caption{Schematic representation of the various regions
of the solar atmosphere and solar wind described
in {\S\S}~4.1--4.4.
\label{fig2}}
\end{figure}

\subsection{Wind Origins in Open Magnetic Regions}

Most of the plasma that eventually becomes the time-steady
solar wind seems to originate in thin magnetic flux tubes
(with observed sizes of order 100--200 km) observed mainly
in the dark lanes between granular cells and concentrated
most densely in the supergranular network.
These strong-field (1--2 kG) flux tubes have been known
as G-band bright points, network bright points, or in
groups as ``solar filigree'' (e.g., Dunn and Zirker 1973;
Spruit 1984; Berger and Title 2001).
Somewhere in the low chromosphere, the thin flux tubes
expand laterally to the point where they merge with one
another into a more-or-less homogeneous network field
distribution of order $\sim$100 G.
At a larger height in the chromosphere, these network
flux bundles are thought to merge again into a large-scale
``canopy'' (Gabriel 1976; Dowdy et al.\  1986).
This second stage of merging is accompanied by further
lateral expansion into ``funnels'' that may be the lowest
sites of observable solar wind acceleration.

SUMER has added substantially to earlier observations of
Doppler shifts and nonthermal line broadening in the
chromosphere, transition region, and low corona
(see H.\  Peter, these proceedings).
Observations of blueshifts in supergranular network
lanes and vertices, especially in coronal holes that host
the fastest solar wind, may be evidence for either the
solar wind itself or upward-going waves that are linked
to wind acceleration processes (e.g., Hassler et al.\  1999;
Peter and Judge 1999; Aiouaz et al.\  2004).
These interpretations are still not definitive, though,
because there are other observational diagnostics that
imply more of a blueshift in the supergranular cell-centers
{\em between} funnels (e.g., He I 10830 {\sc\aa} in coronal
holes; Dupree et al.\  1996; Malanushenko and Jones 2004).

Off-limb measurements with SUMER and CDS have also provided
constraints on plasma temperatures at very low
\linebreak[4]
heights in the corona.
In coronal holes, ion temperatures exceed electron temperatures
even for $r \sim 1.1 \, R_{\odot}$, where densities were
presumed to be so high as to ensure rapid collisional coupling
and thus equal temperatures for all species (Tu et al.\  1998;
Moran 2003).
Spectroscopic evidence is also mounting for the presence
of transverse Alfv\'{e}n waves propagating into the corona
(e.g., Banerjee et al.\  1998).
Electron temperatures derived from line ratios
(David et al.\  1998; Doschek et al.\  2001) exhibit
surprisingly small values in the low off-limb corona
(300,000 to 800,000 K) that are not in agreement with higher
temperatures derived from ``frozen-in'' {\em in situ}
charge states.
The only way to reconcile these observations seems to be
some combination of non-Maxwellian electron velocity
distributions or differential flow between different ion
species even very near the Sun (see Esser and Edgar 2002).

{\em SOHO} has made new inroads into our understanding of
the basic coronal heating problem, but a full review of these
results is is outside the scope of this paper.
The nature of coronal heating must be closely related to the
evolution of the Sun's magnetic field, which changes on
rapid time scales and becomes organized into progressively
smaller stochastic structures (see, e.g.,
Priest and Schrijver 1999; Aschwanden et al.\  2001;
C.\  Schrijver, these proceedings).
As computer power increases, the direct {\em ab initio}
simulation of time-dependent coronal heating is becoming
possible (Gudiksen and Nordlund 2004).

\subsection{The Fast Solar Wind}

It has been known for more than three decades that dark
{\em coronal holes} coincide with regions of open magnetic
field and the highest-speed solar wind streams (Wilcox 1968;
Krieger et al.\  1973).
At the minimum of the Sun's 11-year magnetic cycle the
coronal magnetic field is remarkably axisymmetric (e.g.,
Banaszkiewicz et al.\  1998), with large coronal holes at
the north and south poles giving rise to fast wind
($v_{\infty} > 600$ km/s) that fills most of the heliosphere.
It was fortunate that the first observations of {\em SOHO}
were during this comparatively simple phase, thus minimizing
issues of interpretation for lines of sight passing through
the optically thin outer corona.

One of the most surprising early results from the UVCS
instrument concerned the widths of emission line profiles
of the O~VI 1032, 1037 {\sc\aa} doublet in coronal holes.
These lines were an order of magnitude {\em broader} than
expected, indicating kinetic temperatures exceeding
100 million K at $r > 2 \, R_{\odot}$
(Kohl et al.\  1997).
Because the observational line of sight passes
perpendicularly through the nearly-radial magnetic field
in large coronal holes, this kinetic temperature is a
good proxy for the local ion $T_{\perp}$.
Further analysis of the O$^{5+}$ velocity distribution
was made possible by use of the ``Doppler dimming/pumping''
effect; i.e., by exploiting the sensitivity to the radial
velocity distribution when the coronal scattering profile
is substantially Doppler shifted away from the stationary
profile(s) of solar-disk photons.
This technique allowed the ion temperature anisotropy ratio
$T_{\perp} / T_{\parallel}$ to be constrained
to values of at least 10, and possibly as large as 100.
Temperatures for both O$^{5+}$ and Mg$^{9+}$ were found to be
significantly greater than mass-proportional when compared to
protons (the latter measured by proxy with neutral hydrogen
via H~I Ly$\alpha$ 1216 {\sc\aa}), and outflow speeds for
O$^{5+}$ may exceed those of hydrogen by as much as a
factor of two (see also Kohl et al.\  1998, 1999;
Li et al.\  1998; Cranmer et al.\  1999b).

\begin{figure}[t]
  \begin{center}
    \epsfig{file=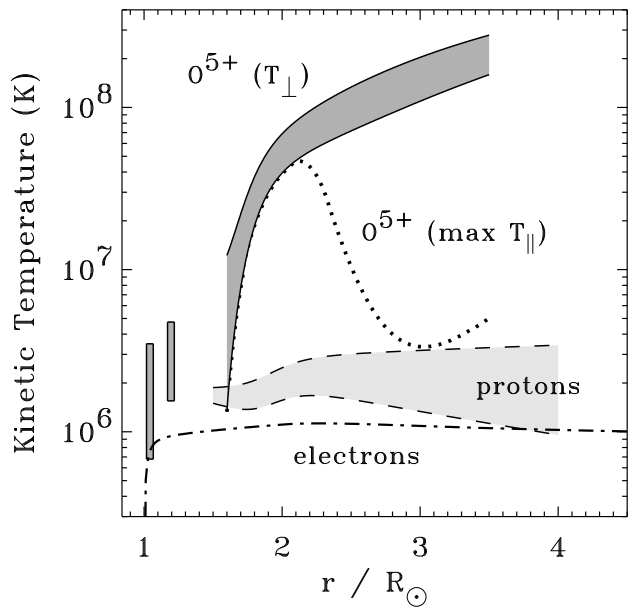, width=8.4cm}
  \end{center}
\caption{Coronal hole kinetic temperatures in the acceleration
region of the fast wind.  Perpendicular temperatures for
protons and O$^{5+}$ above 1.5 $R_{\odot}$ are from an
empirical model that reproduced UVCS data
(Kohl et al., 1998; Cranmer et al., 1999b).
The upper limit on O$^{5+}$ parallel temperature (dotted
line) is from the same empirical model.
The two O$^{5+}$ boxes at lower heights are representative
of ion temperatures derived from SUMER line widths
(Hassler et al., 1997), and the
electron temperature (dot-dashed line) is from an empirically
constrained multi-fluid model (e.g., Hansteen et al.\  1997).
\label{fig3}}
\end{figure}

Figure 3 shows a summary of the solar-minimum coronal hole
temperatures.
Note that when $T_{\rm ion} / T_{p}$ exceeds the mass ratio
$m_{\rm ion} / m_{p}$, it is not possible to interpret the
measurements as a combination of thermal equilibrium and a
spe\-cies-in\-de\-pen\-dent ``nonthermal speed.''
The UVCS and SUMER data (as well as {\em Helios} particle
data at 0.3 AU) have thus been widely interpreted as a truly
``preferential'' heating of heavy ions in the fast solar wind.
Because of the low particle densities in coronal holes, the
observed collection of ion properties has often been
associated with {\em collisionless} wave damping.
The most natural wave mode that may be excited and damped
is the ion cyclotron resonant wave; i.e., an Alfv\'{e}n wave
with a frequency $\omega$ approaching the Larmor frequency of
the ions $\Omega_{\rm ion}$.
The {\em SOHO} observations discussed above have given rise
to a resurgence of interest in ion cyclotron waves
as a potentially important mechanism in the acceleration
region of the fast wind (e.g., McKenzie et al.\  1995;
Tu and Marsch 1995, 1997, 2001; Hollweg 1999, 2000;
Axford et al.\  1999; Cranmer et al.\  1999a;
Li et al.\  1999; Cranmer 2000, 2001, 2002;
Galinsky and Shevchenko 2000; Hollweg and Isenberg 2002;
Vocks and Marsch 2002; Gary et al.\  2003;
Markovskii and Hollweg 2004).

There remains some controversy over whether
ion cyclotron waves generated solely at the coronal base
can heat the extended corona, or if a more gradual and
extended generation of these waves is needed.
If the latter case occurs, there is also uncertainty
concerning the origin of such extended wave generation.
MHD turbulence has long been proposed as a likely means of
transforming fluctuation energy from low frequencies (e.g.,
periods of a few minutes; believed to be emitted copiously
by the Sun) to the high frequencies required by cyclotron
resonance theories (e.g., 10$^2$ to 10$^4$ Hz).
However, both numerical simulations and analytic descriptions
of turbulence indicate that the {\em cascade} from large to
small scales occurs most efficiently for modes that do not
increase in frequency.
In the corona, the expected type of turbulent cascade would
tend to most rapidly increase electron $T_{\parallel}$, not
the ion $T_{\perp}$ as observed.
Cranmer and van Ballegooijen (2003) discussed this issue at
length and surveyed possible solutions.

At times other than solar minimum, coronal holes appear at
all solar latitudes and exhibit a variety of properties.
UVCS has been used to measure the heating and acceleration of
the fast solar wind in a variety of large coronal holes from
1996 to 2004 (Miralles et al.\  2001, 2002;
Poletto et al.\  2002).
A pattern is beginning to emerge, in that coronal holes
with lower densities at a given heliocentric height tend to
exhibit faster ion outflow and higher ion temperatures
(Kohl et al.\  2001).
However, all of the coronal holes observed by both UVCS and
{\em in situ} instruments were found to have roughly similar
outflow speeds and mass fluxes in interplanetary space.
Thus, the densities and ion temperatures measured in the
extended corona seem to be indicators of the range of heights
where the solar wind acceleration takes place.

\subsection{The Slow Solar Wind}

The slow, high-density component of the solar wind is a
turbulent, chaotic plasma that flows at about
300--500 km/s in interplanetary space.
Before the late 1970s, the slow wind was believed to be the
``ambient'' background state of the solar wind, occasionally
punctuated by transient high-speed streams.
This idea came from the limited perspective of spacecraft
that remained in or near the ecliptic plane, and it gradually
became apparent that the fast wind is indeed the more quiet
and basic state (e.g., Feldman et al.\  1976; Axford 1977).

In the corona, the slow wind is believed to originate
mainly from the bright {\em streamers} seen in coronagraph
images.
However, since these structures are thought to be mainly
closed magnetic loops or arcades, it is uncertain how the
wind ``escapes'' into a roughly time-steady flow.
Does the slow wind flow mainly along the open-field edges
of these closed regions, or do the closed fields occasionally
open up and release plasma into the heliosphere?
{\em SOHO} has provided evidence that both processes occur,
but an exact census or mass budget of slow-wind source
regions has not yet been constructed.

\begin{figure}[t]
  \begin{center}
    \epsfig{file=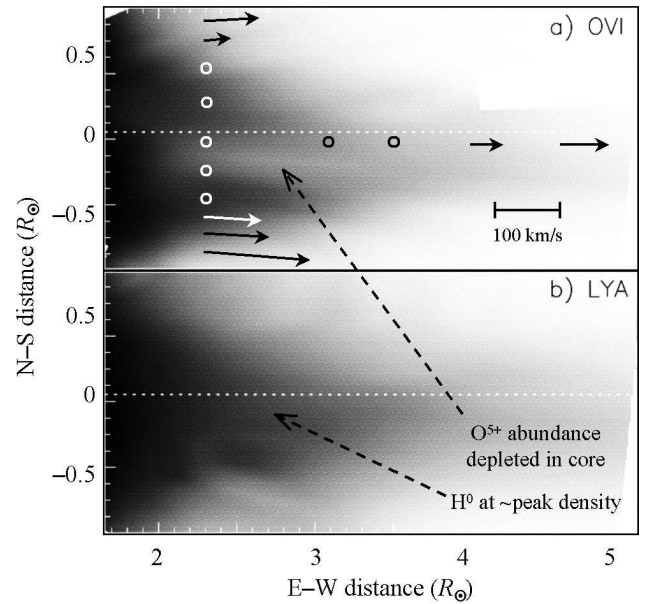, width=8.4cm}
  \end{center}
\caption{Equatorial streamer observed off the west solar limb
with UVCS in April 1997.
Wavelength-integrated intensities are plotted in reverse 
(brightest regions plotted as dark) for
(a) O~VI 1032 {\sc\aa} and (b) H~I Ly$\alpha$.
White dotted lines show the plane of the ecliptic.
Arrows show the computed wind speed, with length proportional
to speed, and circles indicate zero speed (see Strachan
et al.\  2002).
\label{fig4}}
\end{figure}

Figure 4 summarizes several UVCS results concerning
\linebreak[4]
streamers and the slow solar wind.
A comparison of the raster images built up from multiple
scans with the UVCS slit at many heights shows that streamers
appear differently in H~I Ly$\alpha$ and O~VI 1032 {\sc\aa}.
The Ly$\alpha$ intensity pattern is similar to that seen in
LASCO visible-light images; i.e., the streamer is brightest
along its central axis.
In O~VI, though, there is a diminution in the core whose
only interpretation can be a substantial abundance depletion.
At solar minimum, large equatorial streamers showed an oxygen
abundance of 1/3 the photospheric value along the streamer
edges, or ``legs,'' and 1/10 the photospheric value in the core
(Raymond et al.\  1997).
Low FIP (first ionization potential) elements such as Si and
Fe were enhanced by a relative factor of 3 in both cases
(Raymond 1999; see also Uzzo et al.\  2004).
Abundances observed in the legs are consistent
with abundances measured in the slow wind {\em in situ.}
This is a strong indication that the majority of the slow
wind originates along the open-field edges of streamers.
The very low abundances in the streamer core, on the other
hand, are evidence for gravitational settling of the heavy
elements in long-lived closed regions, a result
that was confirmed by SUMER (Feldman et al.\  1998).

Also shown in Figure 4 are outflow speeds derived using the
Doppler dimming method (Strachan et al.\  2002).
Note that a null result (i.e., zero flow speed) was found
at various locations inside in the closed-field core region.
Outflow speeds consistent with the slow solar wind were only
found along the higher-latitude edges and above the
probable location of the magnetic ``cusp'' between about
3.6 and 4.1 $R_{\odot}$.
Frazin et al.\  (2003) used UVCS to determine that O$^{5+}$
ions in the legs of a similar streamer have
significantly higher kinetic temperatures than hydrogen
and exhibit anisotropic velocity distributions with
$T_{\perp} > T_{\parallel}$, much like coronal holes.
However, the oxygen ions in the core exhibit neither this
preferential heating nor the temperature anisotropy.
The analysis of UVCS data has thus led to evidence that
the fast and slow wind share the same physical processes.

Evidence for another kind of slow wind in streamers came
from visible-light coronagraph movies.
The increased photon sensitivity of LASCO over earlier
instruments revealed an almost continual release of
low-contrast density inhomogeneities, or ``blobs,'' from
the cusps of streamers (Sheeley et al.\  1997).
These features are seen to accelerate to speeds of order
300--400 km/s by the time they reach $\sim$30 $R_{\odot}$,
the outer limit of LASCO's field of view.
Wang et al.\  (2000) reviewed three proposed scenarios for
the production of these blobs:
(1) ``streamer evaporation'' as the loop-tops are
heated to the point where magnetic tension is overcome by
high gas pressure;
(2) plasmoid formation as the distended streamer cusp
pinches off the gas above an X-type neutral point; and
(3) reconnection between one leg of the streamer and an
adjacent open field line, transferring some of the trapped
plasma from the former to the latter and allowing it to
escape.
Wang et al.\  (2000) concluded that all three mechanisms
might be acting simultaneously, but the third one seems to
be dominant.
Because of their low contrast, though (i.e., only about
10\% brighter than the rest of the streamer), the blobs
cannot comprise a large fraction of the mass flux of the
slow solar wind.
This is in general agreement with the above abundance
results from UVCS.

Despite these new observational clues, the overall energy
budget in coronal streamers is still not well understood,
nor is their temporal MHD stability.
Recent models run the gamut from simple, but insightful,
analytic studies (Suess and Nerney 2002) to time-dependent
multidimensional simulations (e.g., Wiegelmann et al.\  2000;
Lionello et al.\  2001; Ofman 2004).
Notably, a two-fluid study by Endeve et al.\  (2004) showed
that the stability of streamers may be closely related to the
kinetic partitioning of heat to protons versus electrons.
When the bulk of the heating goes to the protons, the modeled
streamers become unstable to the ejection of massive plasmoids;
when the electrons are heated more strongly, the streamers are
stable.
It is possible that the observed (small) mass fraction of
LASCO blobs can give us an observational ``calibration'' of the
relative amounts of heat deposited in the proton and electron
populations.

Finally, {\em SOHO} has given us a much better means of
answering the larger question:  {\em ``Why is the fast wind
fast, and why is the slow wind slow?''}
A simple, but probably wrong, answer would be that coronal
holes could be heated more strongly than streamers, so it
would be natural for a pressure-driven wind to be accelerated
faster in regions of greater heating.
Even though UVCS has shown that heavy ions in coronal holes
are hotter than in streamers, the proton temperatures between
the two regions may not be that different, and the electrons
are definitely {\em cooler} in coronal holes.

Traditionally, the higher speeds in coronal holes have been
attributed to Alfv\'{e}n wave pressure acceleration;
i.e., the net work done on the plasma by repeated pummeling
from out\-ward-prop\-a\-gat\-ing waves in the inhomogeneous
plasma (e.g., Leer et al.\  1982).
This is likely to be a major contributor, but it does not
address the question of why the wave pressure terms would
be {\em stronger} in coronal holes compared to streamers.
A key empirical clue came from Wang and Sheeley (1990), who
noticed that the eventual wind speed at 1 AU is inversely
correlated with the amount of transverse flux-tube expansion
between the solar surface and the mid-corona.
In other words, the field lines in the central regions of
coronal holes undergo a relatively low degree of
``superradial'' expansion, but the more distorted field lines
at the hole/streamer boundaries undergo more expansion.
Wang and Sheeley (1991) also proposed that the observed
anticorrelation is a natural by-product of {\em equal}
amounts of Alfv\'{e}n wave flux emitted at the bases of
all flux tubes (see also earlier work by
Kovalenko 1978, 1981).

The Wang/Sheeley/Kovalenko hypothesis can be summarized
as follows.  (Any misconceptions are mine!)
In the low corona, the Alfv\'{e}n wave flux $F_A$ is
proportional to
$\rho V_{A} \langle \delta v_{\perp}^{2} \rangle$.
The density dependence in the product of Alfv\'{e}n
speed $V_A$ and the squared wave amplitude
$\langle \delta v_{\perp}^{2} \rangle$ cancels almost
exactly with the linear factor of $\rho$ in the wave flux,
thus leaving $F_A$ proportional only to the radial
magnetic field strength $B$.
The ratio of $F_A$ at the solar wind sonic point
(in the mid-corona) to its value at the photosphere
thus scales as the ratio of $B$ at the sonic point
to its value at the photosphere.
The latter ratio of field strengths is proportional to
$1/f$, where $f$ is the superradial expansion factor as
defined by Wang and Sheeley.
For equal wave fluxes at the photosphere for all regions,
coronal holes (with low $f$) will thus have a larger flux
of Alfv\'{e}n waves at and above the sonic point compared
to streamers (that have high $f$).
In other words, for streamers, more of the energy flux
will have been deposited {\em below} the sonic point.
It is worthwhile to recall that the response of the
solar wind plasma to extended acceleration and heating
depends on whether the energy is deposited in the subsonic
or supersonic wind (Leer and Holzer 1980; Pneuman 1980).
Adding energy in the subsonic, i.e., nearly hydrostatic,
corona raises the density scale height and decreases the
asymptotic outflow speed.
Adding energy above the sonic point results mainly in a
larger outflow speed because the ``supply'' of material
through the sonic point has already been set.
This dichotomy seems to agree with the observed differences
between coronal holes and streamers.

(For other simulations showing how the wind speed depends
on the flux tube divergence, see Chen and Hu 2002;
V\'{a}squez et al.\  2003.
For a recent summary of low-frequency Alfv\'{e}n wave
propagation, reflection, and damping from the photosphere
to the interplanetary medium, see
Cranmer and van Ballegooijen 2004.)

\subsection{Space Weather and CMEs}

In addition to the relatively time-steady solar wind, the
Sun exhibits periodic eruptions of plasma and magnetic
energy in the forms of flares, eruptive prominences, and
coronal mass ejections (CMEs).
These ``space weather'' events have the potential to
interrupt satellite communications, disrupt ground-based
power grids, and threaten the safety of orbiting astronauts
(see, e.g., Feynman and Gabriel 2000; Song et al.\  2001).

{\em SOHO} observations of CMEs have demonstrated this
mission's capability to combine high resolution
imaging with sensitive spectral measurements to obtain
the morphology, evolution, and plasma parameters of the
ejected material.
As the rate of CME events increased from solar minimum to
solar maximum, many unprecedented observations were obtained.
Specifically, EIT, SUMER, and CDS observations contained
information about CME initiation; LASCO constructed a huge
catalog of sizes, morphologies, and expansion speeds of CMEs;
and UVCS provided plasma densities, temperatures,
ionization states, and Doppler shift velocities of dozens
of CMEs in the extended corona (see reviews by
Forbes 2000; Raymond 2002; Webb 2002; Lin et al.\  2003).
UVCS spectra have provided the first real diagnostics of
the physical conditions in CME plasma in the corona, and
they have helped elucidate the roles of shock fronts
(Mancuso and Raymond 2004), thin current sheets driven by
reconnection (Lin et al.\  2004), and helicity conservation
(Ciaravella et al.\  2000).

\section{Conclusions}

{\em SOHO} has made significant progress toward identifying
and characterizing the processes that heat the corona and
accelerate the solar wind.
Most of the {\em SOHO} instruments are expected to continue
performing at full scientific capability for many more
years, hopefully surviving to have some overlap with upcoming
solar missions such as {\em SDO, STE\-RE\-O,} and
{\em Solar-B.}
Unfortunately, none of these missions continue the
UVCS-type coronagraph spectroscopy of the extended corona;
a next-generation instrument of this type would provide
much tighter constraints on, e.g., specific departures
from Max\-well\-i\-an and bi-Max\-well\-i\-an velocity
distributions that would nail down the physics conclusively.
NASA's {\em Solar Probe,} if ever funded fully, would also
make uniquely valuable {\em in situ} measurements of the
solar wind acceleration region.
Observations have guided theorists to a certain extent,
but {\em ab initio} models are still required before
we can claim a full understanding of the physics.
To make further progress, the lines of communication must
be kept open between theorists and observers, and also
between the solar and stellar physics communities.

\begin{acknowledgements}

The author would like to thank John Kohl, John Raymond, and
Andrea Dupree for many valuable discussions.
The groundbreaking {\em SOHO} results would not have been
possible without the tireless efforts of the instrument teams
and operations personnel at the Experimenters' Operations
Facility (EOF) at Goddard.
This work is supported by the National Aeronautics and Space
Administration (NASA) under grants NAG5-11913, NAG5-10996,
NNG\-04G\-E77G, NNG\-04G\-E84G to the Smithsonian Astrophysical
Observatory, by Agenzia Spaz\-i\-ale Italiana, and by the
Swiss contribution to ESA's PRODEX program.

\end{acknowledgements}

\vspace*{-0.1in}
\noindent
\hrulefill

\vspace*{0.1in}
\noindent
{\bf Discussion}

\hspace*{0.1cm}

\noindent
{\em Bob Barber:}
The Sun is a G5 star.  For what other types or classes
of star do you think that your models hold?

\hspace*{0.1cm}

\noindent
{\em Cranmer:}
Late-type stars with hot coronae and solar-type winds
probably extend up the main sequence at least into the
mid-F spectral type and down to M.
Evolved stars between the main sequence and the various
``dividing lines'' in the upper-right H-R diagram also
exhibit coronal signatures and probably have solar-like
mass loss rates (see B.\  Wood, these proceedings).
Even the outer atmospheres of the ``hybrid chromosphere''
stars seem to show some similarities to the solar case.

\hspace*{0.1cm}

\noindent
{\em Andrea Dupree:}
Any time you have strong magnetic fields in a reasonably
high-gravity atmosphere, solar-type coronae and winds
seem to be a natural by-product.
Even a low-gravity supergiant such as Betelgeuse may have
a surface field strength of order 500 G (see B.\  Dorch,
these proceedings) and thus magnetic activity and heating
in its outer atmosphere.

\hspace*{0.1cm}

\noindent
{\em J\"{u}rgen Schmitt:}
You pointed out that the slow wind originates from
individual helmet streamers.  Why is it, then, that the
slow wind at the Earth shows such uniform properties?

\hspace*{0.1cm}

\noindent
{\em Cranmer:}
Well, the slow wind is intrinsically more variable than
the fast wind, but taking this variability into account, it
is true that the slow wind from one streamer looks very much
like the slow wind from another streamer.
This uniformity may be related to the overall uniformity
in the solar wind mass loss rate, which varies only
by about 50\% for {\em all} types of solar wind (e.g.,
Galvin 1998; Wang 1998).
Older solar wind models could not account for this
near-constancy of $\dot{M}$; indeed they predicted that
tiny changes in the coronal temperature would result in
exponentially amplified changes in $\dot{M}$.
Hammer (1982) and Hansteen and Leer (1995) explained this by
modeling the complex {\em negative feedback} that is set up
between thermal conduction, radiative cooling, and
mechanical heating.
This explained the similarity between slow-wind and fast-wind
mass loss rates, so I would assume that it even better
explains the eventual similarity between different source
regions of the slow wind.

\hspace*{0.1cm}

\noindent
{\em Manfred Cuntz:}
Can you comment on the significance of ``polar plumes''
regarding the acceleration of the solar wind?

\hspace*{0.1cm}

\noindent
{\em Cranmer:}
Plumes are bright ray-like features in coronal holes that
seem to trace out the superradial expansion of these
open-field regions.
Plumes are denser and cooler than the ambient ``interplume''
plasma, but there is still some controversy about whether
the solar wind inside them is slower (Giordano et al.\  2000;
Teriaca et al.\  2003) or faster (Gabriel et al.\  2003)
than the flow between plumes.
Wang (1994) presented a model of polar plumes as
the extensions of concentrated bursts of added
coronal heating at the base---presumably via microflare-like
reconnection events in X-ray bright points.
This idea still seems to hold up well in the
post-{\em{SOHO}} era.
EIT and UVCS made observations of compressive MHD waves
channeled along polar plumes (DeForest and Gurman 1998;
Ofman et al.\  1999), and if the oscillations are
slow-mode magnetosonic waves they should steepen into shocks
at relatively low coronal heights (Cuntz and Suess 2001).
Somewhere between about 30 and 100 $R_{\odot}$, plumes seem
to blend in with the interplume plasma, but it is not yet
clear whether this is a gradual approach to transverse
pressure balance or the result of something like a
Kelvin-Helmholtz mixing instability.

\end{document}